# Counting Pairs of Faint Galaxies

David Woods, Gregory G. Fahlman[1] and Harvey B. Richer[1]

Dept. of Geophysics and Astronomy, 2219 Main Mall, University of British Columbia, Vancouver, B.C. V6T 1Z4

astro-ph/9506053    7 Jun 1995

---

[1]Visiting Astronomer, Canada–France–Hawai'i Telescope (CFHT), operated by the National Research Council of Canada, le Centre National de la Recherche Scientifique de France, and the University of Hawai'i.

# ABSTRACT

The number of close pairs of galaxies observed to faint magnitude limits, when compared to nearby samples, determines the interaction or merger rate as a function of redshift. The prevalence of mergers at intermediate redshifts is fundamental to understanding how galaxies evolve and the relative population of galaxy types. Mergers have been used to explain the excess of galaxies in faint blue counts above the numbers expected from no-evolution models. Using deep CFHT ($I \leq 24$) imaging of a "blank" field we find a pair fraction which is consistent with the galaxies in our sample being randomly distributed with no significant excess of "physical" close pairs. This is contrary to the pair fraction of 34% ± 9% found by Burkey *et al.* for similar magnitude limits and using an identical approach to the pair analysis. Various reasons for this discrepancy are discussed. Colors and morphologies of our close pairs are consistent with the bulk of them being random superpositions although, as indicators of interaction, these criteria are uncertain due to contamination of field galaxies and difficulty in seeing structure at intermediate redshifts, respectively. This observed lack of close pairs is probably linked with the decreasing amplitude of the angular correlation function at faint limits. If our faint samples are comprised of galaxies which have a higher average redshift than brighter samples studied by other workers then either the merger rate has been overestimated or there is a change in its behaviour from what is observed at brighter magnitude limits.

*Subject headings:* galaxies–faint, galaxies–pairs

## 1. Introduction

Physically associated pairs of galaxies have long been recognized as harbingers of star formation, AGN behaviour and, in some cases, the inevitable merging of the intially distinct galactic systems. Many theoretical studies have suggested an increasing rate of mergers with redshift to explain the significant evolution of the faint galaxy counts for blue bandpasses (Broadhurst, Ellis and Glazebrook 1992, Carlberg and Charlot 1992 and Colin, Schramm and Peimbert 1994, and references therein). Whether a close pair of faint galaxies is truly in the process of merging can only be determined with redshifts for both objects, and even with this information some assumptions must be made about the critical relative velocity and fraction of close pairs with physical separations larger than those projected on the sky. With a photometric sample of galaxies one must make statistical arguments in order to measure the pair fraction. For faint magnitude limited samples this approach is a necessity since current spectroscopic surveys are limited to $I \sim 22.1$ (Tresse *et al.* 1993 and Lilly *et al.* 1994). Our galaxy sample falls into this category ($I \leq 24$).

The interactions of galaxy pairs in clusters have been shown in recent work by Lavery and Henry (1988, 1994) and Lavery, Pierce and McClure (1992) to be a potentially important mechanism for the "Butcher-Oemler" effect at $z \sim 0.2 - 0.4$. For faint field galaxies the role of interactions and mergers in galaxy evolution are not as well understood or measured. Zepf and Koo (1989, hereafter ZK) compiled a sample of 20 close galaxy pairs from 4m plates of two regions of high galactic latitude down to a magnitude limit of $B_J \leq 22$, for separations less than $4\farcs5$. Comparing to nearby pair samples they found that the frequency of close pairs increases as $(1 + z)^{4.0 \pm 2.5}$. A slight excess of pairs, which may not be statistically significant, was observed for a projected separation of 3". Colless *et al.* (1994) obtained high-resolution imaging of 17 faint blue galaxies culled from their spectroscopic survey (Colless *et al.* 1993) with $z \sim 0.1 - 0.7$ and found that 5 exhibited fainter companions at a projected distance of less than $10h^{-1}$ kpc. Recent HST results (Burkey *et al.* 1994, hereafter BKWF) show 34% of their $I = 18.5 - 23$ galaxies to be close pair members. They claim that this is a lower limit to the pair fraction and find it increases with redshift as $(1 + z)^{3.5 \pm 0.5}$. Since our data is at least one magnitude deeper than that of BKWF we can check the apparently large pair fraction they find and this is our primary motivation in this study. BKWF point out that they detect very few pairs at separations less than $0\rlap{.}''5$, although with HST it is possible to do so. This result suggests that it is feasible to count close pairs from a good ground-based site such as CFHT although the scarcity of sub-arcsecond separation pairs in the field at intermediate redshifts should be confirmed with additional HST data (e.g., Griffiths *et al.* 1994).

Carlberg *et al.* (1994, hereafter CPI) looked at a sample of V magnitude selected ($V \leq 22.5$) galaxy pairs with physical separations less than $\sim 20h^{-1}$ kpc. With redshifts for 14 galaxies in close pairs and 38 field galaxies they found no statistically significant difference between the redshift distributions for the two populations. However, they find an amplitude for the angular correlation function, $\omega(\theta)$, of the field population which is higher, for separations of $\theta \leq 6''$, than an extrapolation of the canonical power-law form $\omega(\theta) \propto \theta^{-0.8}$, as well as a merger rate which goes as $(1 + z)^{3.4 \pm 1.0}$. We study the angular correlation function for faint galaxies in another paper (Woods *et al.* 1995) using a larger galaxy sample but focus on close pairs of galaxies in this work for reasons which are more fully outlined in §2. Griffiths *et al.* (1994) use HST Medium Deep Survey (MDS) data of 201 galaxies with $I \leq 25$, with the caveat that the sample is not complete, to study clustering of faint galaxies and find an excess of nearest neighbours at a projected separation of $\sim 1\rlap{.}''5$. Finally, Yee and Ellingson (1995, hereafter YE) find a similar result to CPI using a magnitude limited sample ($r \leq 21.5$) and some redshifts initially obtained for a quasar-cluster spectroscopic survey. They estimate the fraction of close pairs, with projected separations less than $20h^{-1}$ kpc, to be $\sim 15\%$. Their merger rate of $(1 + z)^{4.0 \pm 1.5}$ agrees with those of CPI, BKWF and ZK (see discussion in §5).

The aim of this study is to determine the pair fraction with V, R and I photometry to fainter limits than previously achieved ($V \leq 25$, $R \leq 24.5$ and $I \leq 24$). We can find close pairs with projected separations as small as ∼ 1" since our summed images in all three colors have subarcsecond seeing. We describe our statistical approach in §2, the processing and analysis of our data are outlined in §3 and the galaxy pairs found are presented in §4. A discussion of our results and conclusions are given in §§5 and 6, respectively.

## 2. Statistical Approach

To determine whether two galaxies are closely aligned on the sky by chance or are a physical close pair, in lieu of redshifts, requires some statistical criteria to attempt to differentiate the two cases. We use BKWF's approach of calculating a statistic which depends on the pair separation and the surface density of galaxies as a function of limiting magnitude. Given a random distribution of galaxies distributed on the sky, the probability of a chance projection occurring for a companion galaxy with apparent magnitude $m$ and separation $\theta$ is:

$$P = \int_0^\theta 2\pi\alpha exp(-\pi\rho\alpha^2)d\alpha = 1 - exp(-\pi\rho\theta^2), \tag{1}$$

with $\rho$ defined as the density of galaxies brighter than $m$. The quantity within the integral sign, the nearest neighbour probability density function, is rigorously derived in an appendix of Scott and Tout (1989). The expression in equation (1) has a correction for both $\omega(\theta)$ and integrals over $\omega(\theta)$, the three-point and higher order correlation functions (White 1979). Since the amplitude of the two-point correlation function is measured to be too small to affect the pair probabilites at our faint magnitude limits (Brainerd et al. 1994), we disregard this correction. The angular two-point correlation function, $\omega(\theta)$, cannot be measured accurately for small separations given the number of galaxies in our one field ($\sim 1000$). Adopting a pair statistic, as above, over correlation analysis is a necessity but it serves our purpose of looking for a significant pair excess at small separations and allows a direct comparison with the BKWF result.

For the $N$ objects detected in each filter to a given magnitude limit, each of the $N(N-1)/2$ possible pairs have their separation calculated along with a local density ($\rho$) calculated by integrating the number counts to the limit of the *faintest* galaxy in the pair. As BKWF point out, integrating to the magnitude of the fainter member is conservative in the sense that the contribution of pair members projected by chance is overestimated. To find close pairs, we adopt a probability of chance projection of $P \leq 0.05$, as did BKWF. The probability cutoff of $P \leq 0.10$ is also used in order to check that the value of P adopted does not have a significant effect on the number of close pairs determined. Since atmospheric seeing effects don't allow us to distinguish pairs to as small a separation as HST we must include a minor correction in equation (1). If $\beta$ is the angular separation cutoff below which individual objects cannot be independently detected then equation 1 becomes:

$$P = \int_{\beta}^{\theta} 2\pi \alpha exp(-\pi \rho \alpha^2) d\alpha = exp(-\pi \rho \beta^2) - exp(-\pi \rho \theta^2). \qquad (2)$$

Adopting a value of $\beta = 0\farcs95$ increases the probability of a chance projection ($P$) by, at most, 0.02 and ultimately adds a few close pairs to our list.

## 3. Photometry Reduction and Analysis

The field we search for close pairs is a high galactic latitude field (hereinafter dubbed "NF1") obtained at the CFHT on 1991 Apr 7-11 totalling 200 min. exposure time in I ($\lambda_{eff}$ = 8320Å, $\Delta\lambda \sim$ 1950Å) and 160 min. in both V ($\lambda_{eff}$ = 5430Å, $\Delta\lambda \sim$ 900Å) and R ($\lambda_{eff}$ = 6485Å, $\Delta\lambda \sim$ 1280Å). Coordinates for the center of the NF1 field are $\alpha(1950) = 13^h 10^m 10\overset{s}{.}80$, $\delta(1950) = +43°01'06\overset{''}{.}0$.

The initial purpose of obtaining this faint field was to look for Population II halo stars (see Richer and Fahlman 1992) but the data is also suitable for studying faint galaxies. The final summed images were obtained from co-adding 1200s frames which were dithered a few arcseconds between exposures. De-biasing and flatfielding were done in the normal manner with a slight variation invoked in the latter process. A median of all the program frames was made to obtain a skyflat for each bandpass. The dithering allowed removal of all, but a couple, of the brightest galaxies and stars to arrive at a smooth final skyflat frame which could be used as an illumination correction frame with the domeflats. To avoid "hotspots" on the final flat-field, a bright galaxy and three saturated stars were masked with the mask being set to the adjacent background level. The individual frames were flat-fielded, registered and summed together to produce a final frame for analysis. The final co-added frames were flattened to better than 1%. Photometric calibration of the galaxies was done with observations of M67 and NGC 4147 standards (Montgomery et al. 1993, Schild 1983 and Christian et al. 1985) taken throughout the run. The seeing FWHM, determined from a number of stars in each summed frame, was found to be $0\overset{''}{.}8$ for I and $0\overset{''}{.}9$ for V and R. The reader is referred to Woods et al. (1995) for a more complete discussion of the data pre-processing, calibration and analysis.

To detect and analyse the galaxies present in the final frames, the FOCAS package was used (Jarvis and Tyson 1981, Valdes 1983 and 1993). The detection threshold is set to 2.5$\sigma$ above the sky with a minimum area of 14 pixels for each object, where $\sigma$ is the dispersion in the sky. A "master" frame (V+R+I) is used to initially detect objects with each bandpass normalized to a common flux level. This technique optimizes the detection of faint objects which typically have a flat spectrum as well as finding galaxies which have extreme colors. The detection algorithm for FOCAS will find different numbers of objects, by a few percent, depending on the orientation of the frame (D. Hudon, priv. communication). This is due to the line-by-line nature of the detection algorithm and the threshold for a particular line depending on the sky history from previous lines. We get around this by rotating the master frame through 90 degree increments and matching the resultant four catalogs to produce a master catalog. Before matching these catalogs, they are filtered to only include galaxies, stars or intermediate objects using the default FOCAS classification scheme. This filtering removes spurious detections such as cosmic rays and saturation artifacts. The master catalog is also filtered to remove detections of objects which lie within "masked" areas of the frame. Masked regions are typically in the vicinity of saturated stars, bad columns or vignetted corners. To ensure the efficacy of the detection algorithm each final sample of galaxies was checked by eye.

The final master catalog is used as a template for the V, R and I summed images to evaluate magnitudes for each bandpass. A $\sim 3"$ diameter aperture is used to measure the magnitudes unless the object in question possesses a characteristic size which exceeds the aperture size, in which case an isophotal magnitude is determined. For some of the pairs with small separations, the magnitudes of the constituent objects will be overestimated due to contamination from the neighbour's flux. This boosting of magnitudes, due to the $3''$ aperture, will be most important for the fainter and smaller galaxies which need to be at smaller angular separations to be counted as a close pair. With overestimated magnitudes a close pair is more apt to be considered real (e.g., $P \leq 0.05$) since the galaxies can have larger separations. This will increase the number of close pairs found but it does not seem to be a significant effect, as shown by the results in §4.

Since NF1 is a high galactic latitude field ($b \simeq 74°$) there is not significant contamination from stars in our galaxy samples. Star-galaxy separation is more of a concern for BKWF since the line of sight to their fields is closer to the galactic plane ($b \simeq 38°$) although their discrimination should be fairly unambiguous. Fig.1 shows the crude shape parameter we use to remove the brighter stars, in this example for the I data. The difference between the "core" magnitude, corresponding to the central 3x3 pixels of each object, and the aperture magnitude is plotted versus the aperture magnitude. A clear stellar sequence is observed at $(m_{core} - m_{aper}) \sim 1.1$ with the shape parameter rising to larger values for bright magnitudes due to saturation. Using this crude discriminant we can separate the stars and galaxies down to $I \sim 21$, $V \sim 22$ and $R \sim 21$. The identification of the stars are double checked with the more sophisticated star-galaxy separation technique employed by two of the authors (GGF and HBR) to identify Population II halo stars. Fainter than the limits given above the stars are left in the sample since their contribution is negligible at best and we want to avoid removing compact galaxies which have a stellar appearance. From the halo star study it is estimated that there are 38 stars for the magnitude range $21 < I < 24$ leaving 869 galaxies with this brightness. Regardless of whether the stars are included in the final samples or not, there is not a significant effect observed on the determined close pair fractions.

Number counts determined for the three bandpasses are given in Fig.2. Magnitude limits are conservatively estimated to be $V \simeq 25$, $R \simeq 24.5$ and $I \simeq 24$. These counts are used to determine the density of objects as a function of magnitude ($\rho(m)$) required for the pair statistic. Note that the V and I counts have been shifted along the abscissa for illustrative purposes. The dashed line is a least squares fit to the counts, using data points with brighter magnitudes than the magnitude limits, for each color. Slopes calculated for the V, R and I number counts are $0.41 \pm 0.01$, $0.36 \pm 0.01$ and $0.32 \pm 0.01$, respectively. These values agree well with those published in the literature. For the R and I counts Tyson (1988) finds a slope of 0.39 and 0.34, while Lilly *et al.* (1991) obtain 0.32 for I. We define a "bright" sample with magnitudes of: $19 < V < 24$, $18.5 < R < 23.5$ and $18 < I < 23$ with 359, 391 and 496 galaxies, respectively. Our "faint" sample goes one magnitude deeper in each color, to the completeness limits, resulting in 938 galaxies in V, 891 in R and 1005 in I. These six photometry samples were subjected to the pair statistical analysis. The data from each bandpass will have a similar limit in magnitude since $I \sim V + 1$ and $I \sim R + 0.5$ for galaxies with late-type morphologies out to $z \simeq 0.4$ (Frei and Gunn 1994). Exceptions to this are ellipticals which become harder to detect since the 4000Å break has been redshifted beyond the V filter at $z \geq 0.4$.

## 4. Galaxy Pairs

Using the technique outlined in §2, we find all the close pairs in the bright and faint V, R and I samples for probability cutoffs $P \leq 0.05$ and $P \leq 0.10$ and separations ranging from $1 - 11''$. A physical separation of 20 $h^{-1}kpc$, using a Hubble constant of $H_0 = 100\ km\ s^{-1}Mpc^{-1}$ ($h = 1$) and $q_0 = 0$, for redshifts of 0.2 and 0.8 corresponds to angular separations of $9''$ and $4''$, respectively. Therefore a considerable range of angular separations between pairs is studied although at small separations we are limited at $1''$ (corresponds to $\sim 4$ kpc at $z = 0.5$) by seeing effects. Our results are shown in Figs. 3 and 4 for the bright and faint samples respectively. The solid-lined bar histograms are the observed close pairs which satisfy the statistical criterion. The solid line curves are the distributions of pairs expected if the galaxies were randomly distributed in the images. In each case, these were calculated from the average of a pair analysis of 1000 catalogues constructed by assigning the observed galaxies random coordinates in the field. The dashed lines above and below the solid line represent the 95% confidence levels bracketing the expected random pair distribution.

One complication we do not consider in the random simulations is the angular extent of the galaxies. This shouldn't be important since the galaxies in our images have small average sizes and the total area of the images covered by galaxies is a few percent of the total field. In these random simulations of galaxies distributed in our field any "pairs" found with separations $< 1''$ are considered to be one object since this is the small separation cutoff we use in the real data, as is evident from the histograms in Figs. 3 and 4. In the simulations objects are not allowed to occupy the positions corresponding to the masked regions in the observed frames. Another potential problem with our image analysis could be that FOCAS has difficulty splitting objects which have small angular separations. This was manually checked by looking at all the objects detected in the frames over the full magnitude range and no evidence was found for a significant number of close pairs being missed by the detection algorithm. In particular, for the bright sample to $I \sim 23$ where we make the direct comparison with BKWF, this is definitely not a problem.

Note that all the observed histograms, in Figs. 3 and 4, are *consistent with a random distribution of galaxies*, with the possible exception of the bright samples of galaxies with $P \leq 0.10$. A slight excess of close pairs is observed in some of these cases for separations around $3 - 5''$, albeit at the $\sim 2\sigma$ level. The pair fractions determined in each sample for the separation range of $1 - 11''$ are listed in Tables 1 and 2, along with the expected fractions for a random distribution of galaxies determined from our Monte Carlo simulations. Errors for the pair fractions ($1\sigma$) are calculated from binomial statistics, while errors for the random distribution fractions are not given since they can be made arbitrarily small, in principle, if one runs enough simulations. For the conservative probability cutoff of $P \leq 0.05$ all our pair fractions agree with the random values, for both the bright and faint samples in all three bandpasses, within $\sim 2\sigma$ errors. With the more liberal probability cutoff $P \leq 0.10$ the V and R samples have pair fractions which depart by more than $3\sigma$ from that expected for a purely random distribution of galaxies but only the bright V sample has a significant excess.

This is contrary to what BKWF found with a sample of galaxies ($18.5 < I < 23$) similar to our bright I sample, using an identical approach considering pairs with $P \leq 0.05$, where they measured a pair fraction of 34% ± 9%. The pair fraction we measure is 10.1 ± 1.4% but this becomes consistent with a zero fraction of physical pairs after the non-physical pairs expected from random superpositions (9.4%) are subtracted. One of the principal reasons for the discrepancy with BKWF is that they did not correct their pair fraction for randomly distributed galaxies. YE have also pointed this out and after making a correction for optical pairs claim that the *physical* pair fraction in the BKWF data is $\sim 17\%$, which agrees with CPI and their value. If the physical pair fraction at $I \sim 23$ was $\sim 17\%$ we should have had no problem detecting these close pairs in our images. In the only case where a significant excess above the random pairs expectation is observed, for the bright V sample with $P \leq 0.10$, a physical pair fraction of 10.4 ± 3.1% is implied. This is not a particularly large excess considering the BKWF result and since we have doubled their probability cutoff, to 0.1, to obtain it. Therefore, our main result is that *there is no significant excess of close pairs for the various magnitude limited samples we consider.*

Another reason why the pair fraction obtained by BKWF may have been higher than our result is that they do not use a uniform magnitude limit for their pair analysis. From their number counts they appear to be complete down to $I \sim 22.3$ but it is mentioned that there are differing magnitude limits in the separate fields studied. Presumably surface brightness selection effects should roughly be the same for single and paired galaxies, as discussed by BKWF. However, to calculate accurate pair fractions within a magnitude range it is important to have complete data so the total number of galaxies is well measured. We are confident that our data is complete to the limits ($I \sim 23$) that BKWF quote their pair fractions for. Also, a concern with both the BKWF and YE studies is the choice of the fields used for finding close pairs. BKWF have fields centered on or near faint radio galaxies which they claim are not in rich or poor clusters. YE obtain a sample of field galaxies from their quasar-cluster redshift survey by using an absolute velocity difference between the galaxies and the quasar in each CCD frame of greater than 4000 $km\ s^{-1}$. Galaxies fainter than $r \leq 21.5$ are not considered due to the success rate of determining redshifts being less than 78% below this magnitude limit. Our data is of a more pristine nature for pair studies since the NF1 blank field was specifically chosen to avoid Zwicky galaxy clusters and any other evidence of clustering.

To illustrate the close pairs we detect from our samples, a mosaic of images of pairs found for the sample which is comparable to that of BKWF ($P \leq 0.05$, $18 < I < 23$) is presented in Fig. 5. Each tickmark on the axes corresponds to $1''$. Clearly the fainter galaxies must be observed at smaller separations to be considered members of close pairs. Table 3 lists the magnitude of each object which is in a close pair in the corresponding lettered images. The galaxies are numbered and the magnitudes are given from left to right in each image. Calculated *a posteriori* probabilities (P) are given for each pair in the third column. In cases where there is more than one close pair in the frame the probabilities listed are for the galaxies given in the following brackets.

Morphological peculiarities such as tidal tails and distortions of the isophotes of the galaxies are typically used as indicators of an interaction or merging event. With increasing redshift and decreasing resolution these subtle, low surface brightness features are difficult to detect, let alone quantify. Using simulations of WFPC2 HST images of galaxy mergers at $z = 0.4$ and $z = 1.0$, Mihos (1995) demonstrates that using these morphological signatures is subject to significant uncertainty due to the rapid evolution of the interacting system once the galaxies have merged. With large redshifts or poorer resolution (ground-based) images he suggests that merging systems can only be found through the presence of companions. As a test, we looked for morphological signatures of interactions or mergers in our deepest image for the close pairs found with $P \leq 0.10$ and $18 < I < 23$. Only $\sim 7 - 10$ out of 66 close pairs showed unambiguous peculiarities indicative of tidal perturbation. The identification of these features is limited by the seeing and pixelization effects and this identified fraction is certainly a lower limit. Nevertheless, the relatively small number of objects with perturbed morphology agrees with our assertion that the majority of the close pairs found by the statistical methodology applied here are merely chance alignments.

Another test of the physical nature of our pair samples is to determine their color distributions. Galaxies which are undergoing interactions or mergers will have starbursts induced subsequently causing the colors to be bluer (Larson and Tinsley 1978). In Fig. 6 we plot the colors for our bright samples in each bandpass. Only the bright samples are plotted since we can obtain the most accurate colors for these objects and also, due to these galaxies having the only significant pair excess above that expected for non-physical pairs. Colors are determined using 3″ aperture magnitudes for all galaxies. The solid-line histograms are the color distributions of the bright sample detected in that filter, and the dashed and dotted lines correspond to the colors of galaxies occupying close pairs for $P \leq 0.05$ and $P \leq 0.10$, respectively. No K-corrections are made since the mean redshifts of the entire galaxy and pair distributions should be basically the same (CPI). A $\chi^2$ test shows no significant differences between the general galaxy color distributions and that of the pairs, with either $P \leq 0.05$ and $P \leq 0.10$, for all three bandpasses. Therefore, one cannot rule out the pair and total galaxy distributions being drawn from the same parent population. This also supports our conclusion that the majority of close pairs we have identified are optical pairs and not physically associated. However, as YE comment, separating the true interacting systems by color, from those which are either optically aligned or physical pairs with high relative velocities, is very difficult given the small ratio of the former to the latter. Also with the smaller separation pairs, where we would expect at least one galaxy in the pair to have bluer colors than the general distribution, there is going to be more flux from the neighbouring object contaminating the aperture for which the color measurement is made.

## 5. Discussion

Whether merging increases the number of close pairs or not depends on the merger timescale and when galaxies brighten during the interaction process (Broadhurst *et al.* 1992). Toomre (1977) has used the occurrence of galaxy pairs with tidal tails in a sample of $\sim 4000$ RC2 galaxies to estimate the time for two local galaxies to merge to be $\sim 0.5$ Gyr. With the stringent requirement of observable tails this is probably a conservative number and CPI, using more elaborate arguments, derive a local merging timescale of $\sim 22 \pm 2.6$ Gyr. At higher redshifts CPI assume the pairwise velocity dispersion evolves as $(1+z)^{-1}$ yielding a merger timescale of $7.1 \pm 1.4$ Gyr at $z = 0.4$.

There is general agreement that the measured local pair fraction is $\sim 4 - 5\%$ (Soares *et al.* 1995, CPI, YE). The determinations of the merging rate typically parameterize it as a power law $(1+z)^m$, where $m$ has been estimated to lie within the range of $\sim 2.5 - 4$ with substantial errors associated with the exponent (Toomre 1977, ZK, CPI, BKWF and YE). The pair fraction growth can also be expressed in the form $(1+z)^n$. For this parameterization, BKWF assume that the merger rate evolves as $(1+z)^{n-1}$ while CPI use dynamical reasoning to obtain a merger rate which increases as $(1+z)^{n+1}$. Finally, YE suggest that the observed pair fraction is a good estimator of the actual merging population at a given redshift so that the merging rate is $(1+z)^n$ (i.e., $m \simeq n$). For the $I \sim 23$ limit YE find the physical BKWF close pair fraction to be $\sim 17\%$ when the optical pairs are accounted for. If this pair fraction exists at $I \sim 23$, or even to $I \sim 24$, we should be able to measure it with our photometry since we are using an identical approach to the analysis as BKWF.

So, why do we not observe a strong excess in the pair fraction for our magnitude limits? It should be noted that the magnitude limits for which CPI ($V \sim 22.5$) and YE ($r \sim 21.5$) calculate their pair fractions are much brighter than our limits and BKWF's. Assuming the average redshift of galaxies in our bright and faint samples are higher than that of the CPI and YE samples, there should be an increase in the pair fraction above the $\sim 10 - 15.5\%$ they measure. The fact that we only see a comparable pair fraction for the bright V sample with $P \leq 0.10$, while the other samples have substantially lower fractions, suggests that either the exponent for the merging rate has been overestimated or there is a dropoff in mergers at our magnitude limits. Another possible complication is that there may be different merger behaviour depending on the dominant morphological type of galaxy at the specific magnitude limit (Glazebrook *et al.* 1994, Driver *et al.* 1994). Ideally the pair fraction evolution should be determined as a function of galaxy morphology and magnitude.

We can consider the angular correlation function since it is related to the expected pair fraction, although measuring $\omega(\theta)$ at small scales with small samples of galaxies has inherently large errors. As an additional test, we measured $\omega(\theta)$ for the bright and faint magnitude limits in the V, R and I images down to a separation of $2''$. The $\omega(\theta)$ calculated for all cases were found to be consistent with randomly distributed galaxies within the substantial error limits. This is what is expected from the pair fractions we determine. A better determined $\omega(\theta)$ from multiple fields with larger galaxy samples will be discussed in Woods *et al.* (1995).

Brainerd *et al.* (1994) have measured $\omega(\theta)$ to the faintest limits to date with a sample of $\sim 5700$ galaxies to $r \sim 26$, which corresponds to $R \sim 25.5$, down to separations of $\sim 22''$. Since the amplitude of the correlation function is observed to fall off monotonically with the magnitude limit of the sample, we use the Brainerd *et al.* fits and their Fig. 2 to estimate the amplitude of $\omega(\theta)$ for our bright and faint magnitude limits of $R \sim 23.5$ and $R \sim 24.5$. A form for the correlation function of $\omega(\theta) = A_\omega \theta^{-0.8}$ has been adopted. By extrapolating $\omega(\theta)$ to small scales a prediction for the fraction of "non-random" pairs (those *not* resulting from random superpositions), expected for a particular angular separation, can be made. For the bright limit in R this gives $\omega(1'') \simeq 0.6$ and $\omega(1'') \simeq 0.31$ for the faint sample. These values correspond to $\sim 38\%$ and $\sim 24\%$ of the pairs being non-random. If we use CPI's operational definition which requires a physical pair to have a separation $\theta$ such that $\omega(\theta) \geq 1$, then our estimated values for $\omega(\theta)$ do not bode well for finding "real" pairs. If the extrapolation of the Brainerd *et al.* $\omega(\theta)$ to small separations is to be believed and since our minimum pair separation is $1''$, it is not surprising we do not detect an excess of close pairs, at least in our R data. So, it may be a consequence of the weak amplitude of the angular correlation function at these magnitude limits that a stronger pair fraction is not observed in our sample, although we emphasize that a measurement of $\omega(\theta)$ is not being made but an extrapolation. More accurate determinations of $\omega(\theta)$, using larger numbers of galaxies than previous samples, down to faint limits and small angular separations with good angular resolution are needed. New CCD mosaic cameras with wide-field imaging capabilities currently becoming available would be ideal for this task and these data would further constrain merger models (see, for e.g., Fig. 5 in Carlberg and Charlot 1992). It is interesting to note that the MDS survey, using HST, has determined $\omega(\theta)$ down to $I \sim 22$ for small separations (a few $''$) and found no significant excess of close pairs above a canonical power law slope (Neuschafer *et al.* 1995).

## 6. Conclusions

Our principal result is that we find no evidence for a significant excess of close pairs of galaxies for magnitude limits of $I \sim 23$ and $I \sim 24$ (and similarly for $R \sim 23.5, 24.5$; $V \sim 24, 25$). This result is contrary to the large pair fraction found by BKWF for $I \leq 23$. If the probability cutoff $P \leq 0.10$ is used slight pair excesses are found in our V and R samples although typically yielding substantially smaller pair fractions than those observed by other workers at brighter limits. The lack of a large physical pair fraction limits the usefulness of small photometric surveys as a tool for studying the merger rate among the faint galaxies. Further, if the faint galaxies studied here are representative of a more distant sample than the galaxy samples of CPI and YE then this suggests the merger rate has either been previously overestimated or there is a change in its behaviour beyond their brighter magnitude limits. Of course, fainter galaxy samples ($I \sim 23$ or 24) need not have higher average redshifts than bright samples since spectroscopic surveys haven't measured redshift distributions for these faint limits. The relative contribution of different galaxy morphologies to the number counts at a given magnitude limit (Driver *et al.* 1994) could also affect the observed merger rate for the total galaxy sample. The absence of a significant excess of galaxy pairs is consistent with an *extrapolation* of the angular correlation function for faint galaxies to smaller separations (Brainerd *et al.* 1994). This conclusion should be verified by a direct measurement of $\omega(\theta)$ for $\theta \simeq 1'' - 10''$.

Additional tests of looking for morphological signatures of interactions or differences between the colors of close pairs and the total galaxy population agree with our conclusion that the majority of our pairs are chance alignments, with the caveat that these criteria are subject to considerable uncertainties. Our result doesn't preclude there being *any* mergers at our magnitude limits (some obviously interacting galaxies are observed in our data) but suggests that there are fewer close pairs than what is observed with brighter samples. Although our CCD imaging has better than $1''$ seeing, additional HST data of appropriate "blank" regions with faint magnitude limits is required to study the frequency of merging systems in the field with sub-arcsecond separations. To unambiguously determine the physical pair fraction, and thereby the merger rate, as a function of lookback time many more redshifts of faint, isolated and paired galaxies are required.

– 19 –

We thank the CFHT TAC for a generous allocation of observing time. For several helpful, interesting discussions and for making preprints promptly available, we would like to thank Ray Carlberg, Dan Hudon, Frank Valdes and Howard Yee. Support for this research was made possible through grants to GGF and HBR from the National Science and Engineering Research Council of Canada.

– 20 –| Probability | $18 < I < 23$ | Random | $18.5 < R < 23.5$ | Random | $19 < V < 24$ | Random |
|---|---|---|---|---|---|---|
| $P \leq 0.05$ | $10.1 \pm 1.4\%$ | $9.4\%$ | $11.3 \pm 1.6\%$ | $9.2\%$ | $13.6 \pm 1.8\%$ | $9.3\%$ |
| $P \leq 0.10$ | $22.0 \pm 1.9\%$ | $17.8\%$ | $24.3 \pm 2.2\%$ | $17.6\%$ | $27.9 \pm 2.4\%$ | $17.5\%$ |

Table 1: Pair fractions for the bright samples in V, R and I for both probability cutoffs.

| Probability | $18 < I < 24$ | Random | $18.5 < R < 24.5$ | Random | $19 < V < 25$ | Random |
|---|---|---|---|---|---|---|
| $P \leq 0.05$ | $8.7 \pm 0.9\%$ | $9.4\%$ | $10.2 \pm 1.0\%$ | $9.4\%$ | $12.0 \pm 1.1\%$ | $9.4\%$ |
| $P \leq 0.10$ | $19.8 \pm 1.3\%$ | $18.2\%$ | $23.0 \pm 1.4\%$ | $18.1\%$ | $22.7 \pm 1.4\%$ | $18.1\%$ |

Table 2: Pair fractions for the faint samples in V, R and I for both probability cutoffs.

– 21 –| Image (Fig.5) | I Magnitudes | P |
|---|---|---|
| a | 22.67, 20.88, 21.09 | 0.037(1,2), 0.003(2,3) |
| b | 21.06, 21.34 | 0.008[a] |
| c | 19.83, 20.44 | 0.031 |
| d | 20.96, 19.58 | 0.009 |
| e | 21.95, 22.38 | 0.007 |
| f | 21.73, 22.69 | 0.020 |
| g | 22.39, 22.74 | 0.035 |
| h | 21.09, 22.28 | 0.010 |
| i | 22.89, 21.09 | 0.042 |
| j | 19.83, 20.03, 19.85 | 0.021(1,2), 0.048(1,3), 0.010(2,3) |
| k | 22.57, 20.72 | 0.020 |
| l | 22.37, 22.59 | 0.033 |

Table 3: Listing of probabilities of galaxy pairs in Fig.5.

[a]The brighter object on the left has a radial profile consistent with that of a star. It hasn't been removed since it is fainter than the $I \sim 21$ cutoff below which no attempt is made to remove stellar objects.

| Image (Fig.5) | I Magnitudes | P |
|---|---|---|
| m | 20.99, 21.28 | 0.006 |
| n | 20.55, 22.14 | 0.025 |
| o | 20.18, 21.41, 20.90 | 0.041(1,2), 0.030(2,3) |
| p | 20.45, 20.83 | 0.033[b] |
| q | 22.69, 22.14 | 0.017 |
| r | 21.93, 22.51 | 0.010 |
| s | 22.49, 22.73, 21.94 | 0.040(1,3), 0.017(2,3) |
| t | 22.42, 22.95 | 0.048 |
| u | 22.55, 22.10 | 0.022 |
| v | 19.01, 18.43 | 0.006 |
| w | 18.41, 20.12 | 0.048 |

[b]The galaxy on the left is actually a close pair missed by the detection algorithm. Galaxies are easily detected in pairs down to angular separations of 1", so, this failure is a rare occurrence.

---

Fig. 1.— Our crude concentration parameter, $m_{core} - m_{aper}$, is plotted vs. the aperture magnitudes in the I filter. Note the stellar sequence which is apparent down to $I \sim 21$. The magnitude limit is estimated to be $I \sim 24$.

Fig. 2.— Number counts for the V, R and I bandpasses are shown for our sample. The lines are least squares fits to all the points above the magnitude limits of $I \leq 24$, $R \leq 24.5$ and $V \leq 25$. Slopes are determined to be $0.41 \pm 0.01$, $0.36 \pm 0.01$ and $0.32 \pm 0.01$, respectively. V counts have been shifted to the right and I counts to the left, by 2 magnitudes, for purposes of clarity.

Fig. 3.— Observed close pair histograms for the "bright" magnitude limited samples in V, R and I, for probability cutoffs of $P \leq 0.05$ and $P \leq 0.10$. The solid curves are the expected pairs for galaxies which are distributed randomly in our images. Dashed lines correspond to the 95% confidence levels above and below the average random distribution, which have all been determined from 1000 simulations.

Fig. 4.— Same as Fig.3 except these histograms are for the "faint" magnitude limited samples in V, R and I. Note the different scale in the number coordinate from Fig.3.

Fig. 5.— Images of the galaxy pairs found for $18 < I < 23$ and an *a posteriori* probability of $P \leq 0.05$. Magnitudes and calculated probabilities for each galaxy pair are listed in Table 3 for the objects shown from left to right.

Fig. 6.— Histograms of the color distributions of pair samples found for the "bright" samples of galaxies in each bandpass with $19 < V < 25$, $18.5 < R < 24.5$ and $18 < I < 24$, compared to the colors of all the galaxies in these magnitude limits. The solid line histogram is the general galaxy distribution while the dashed and dotted lines correspond to the pairs found with $P \leq 0.05$ and $P \leq 0.10$, respectively. All colors are determined using $3''$ aperture magnitudes.

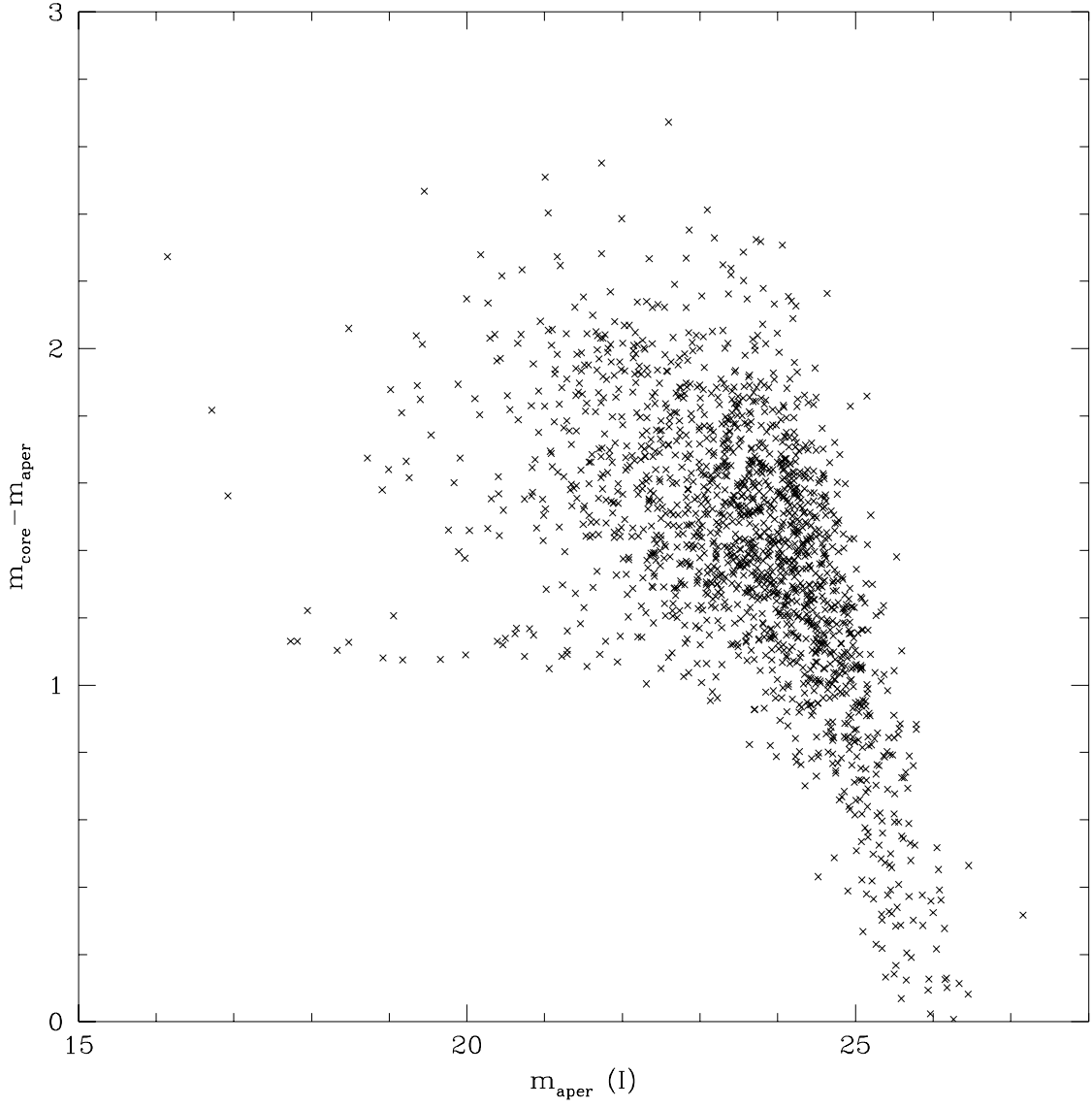

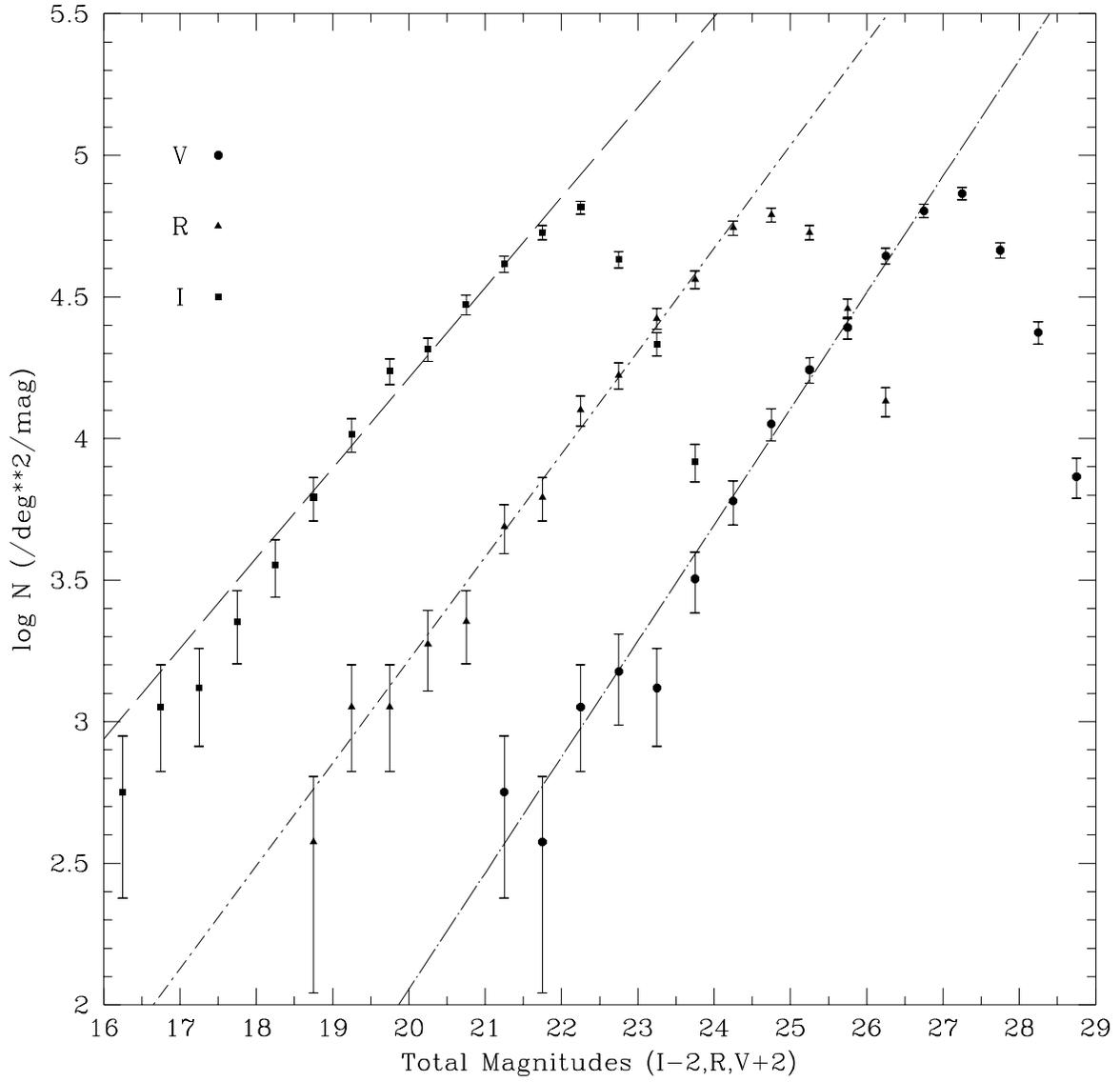

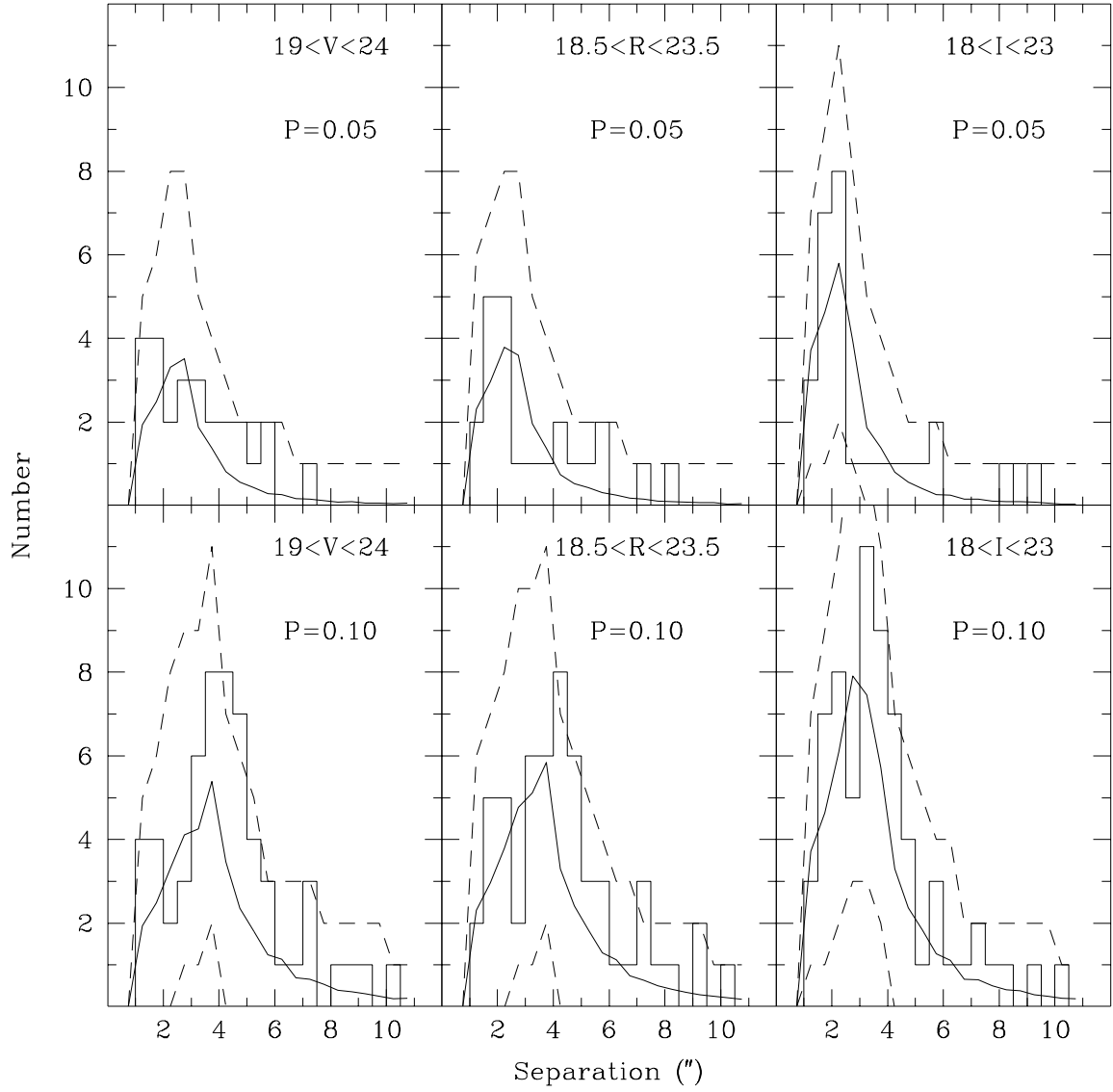

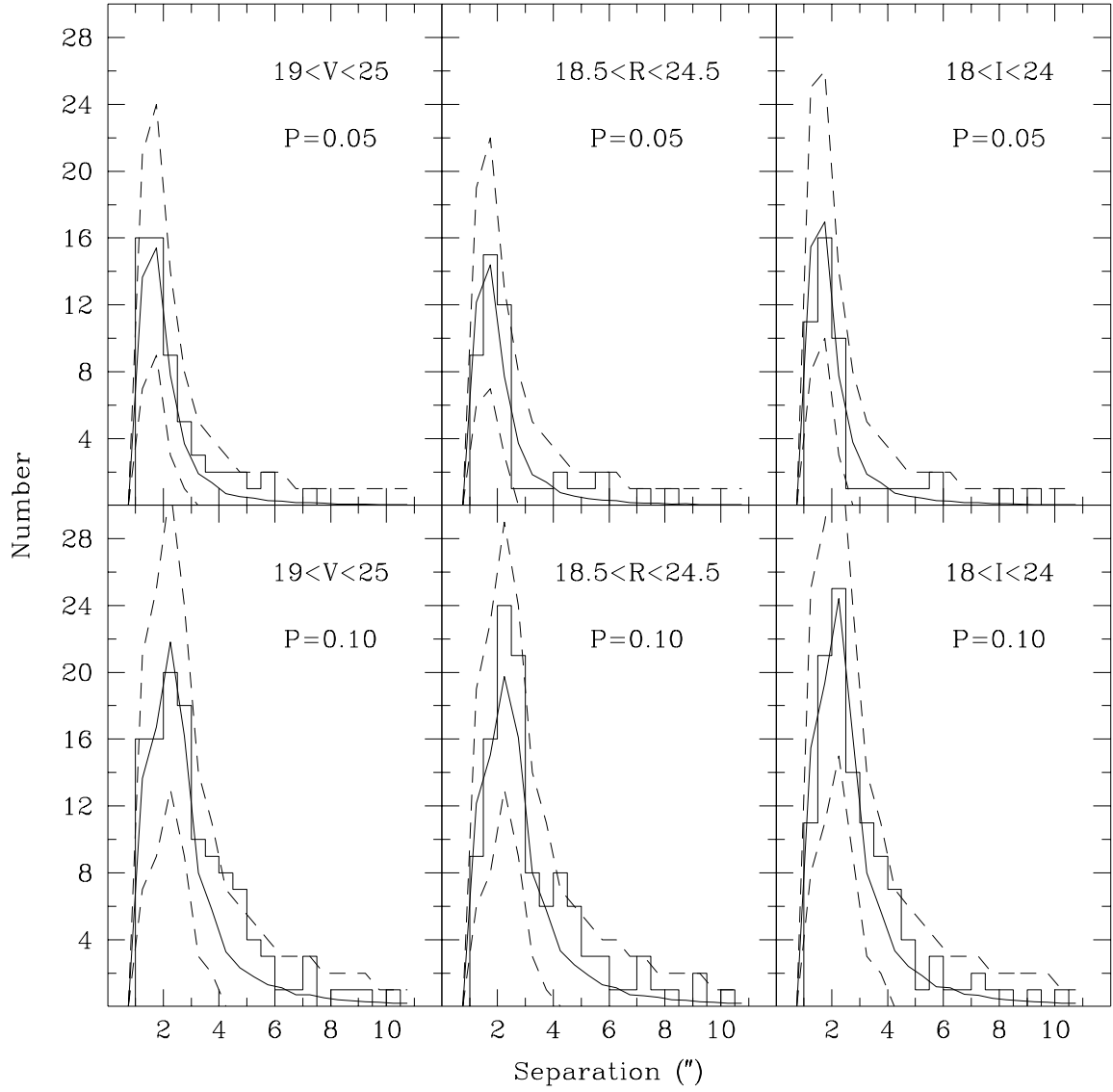

*Figure 5*

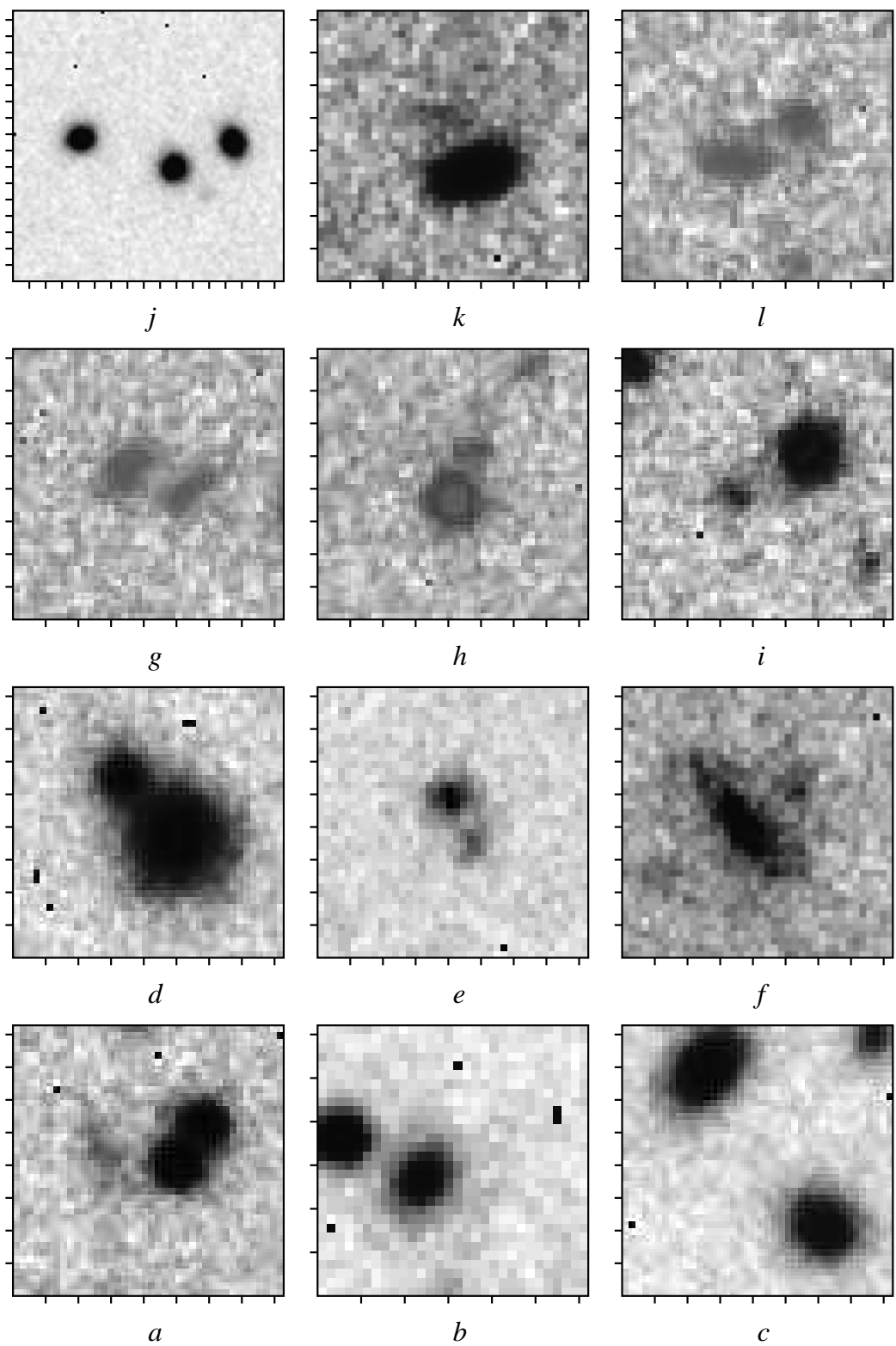

*Figure 5*

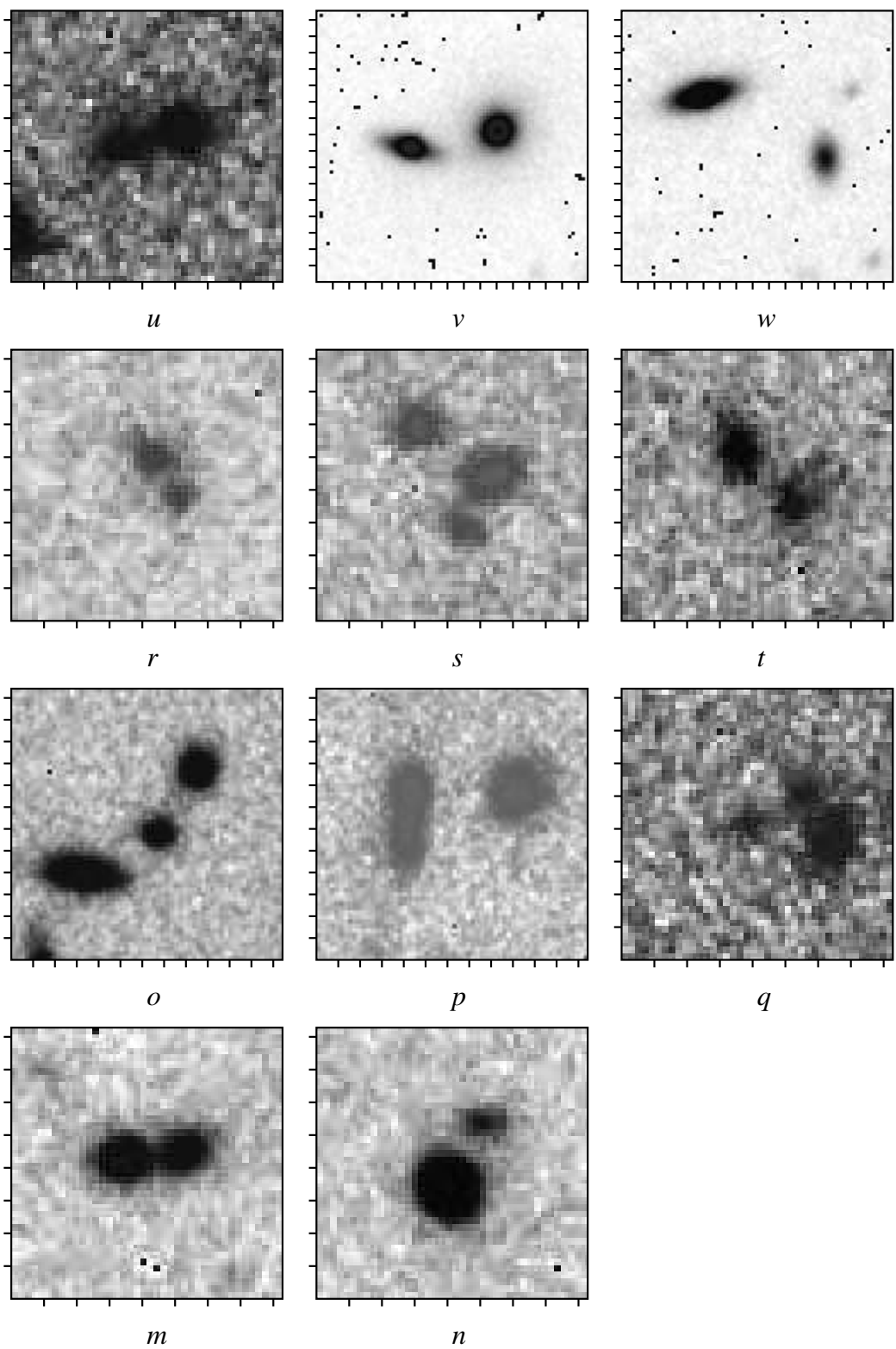

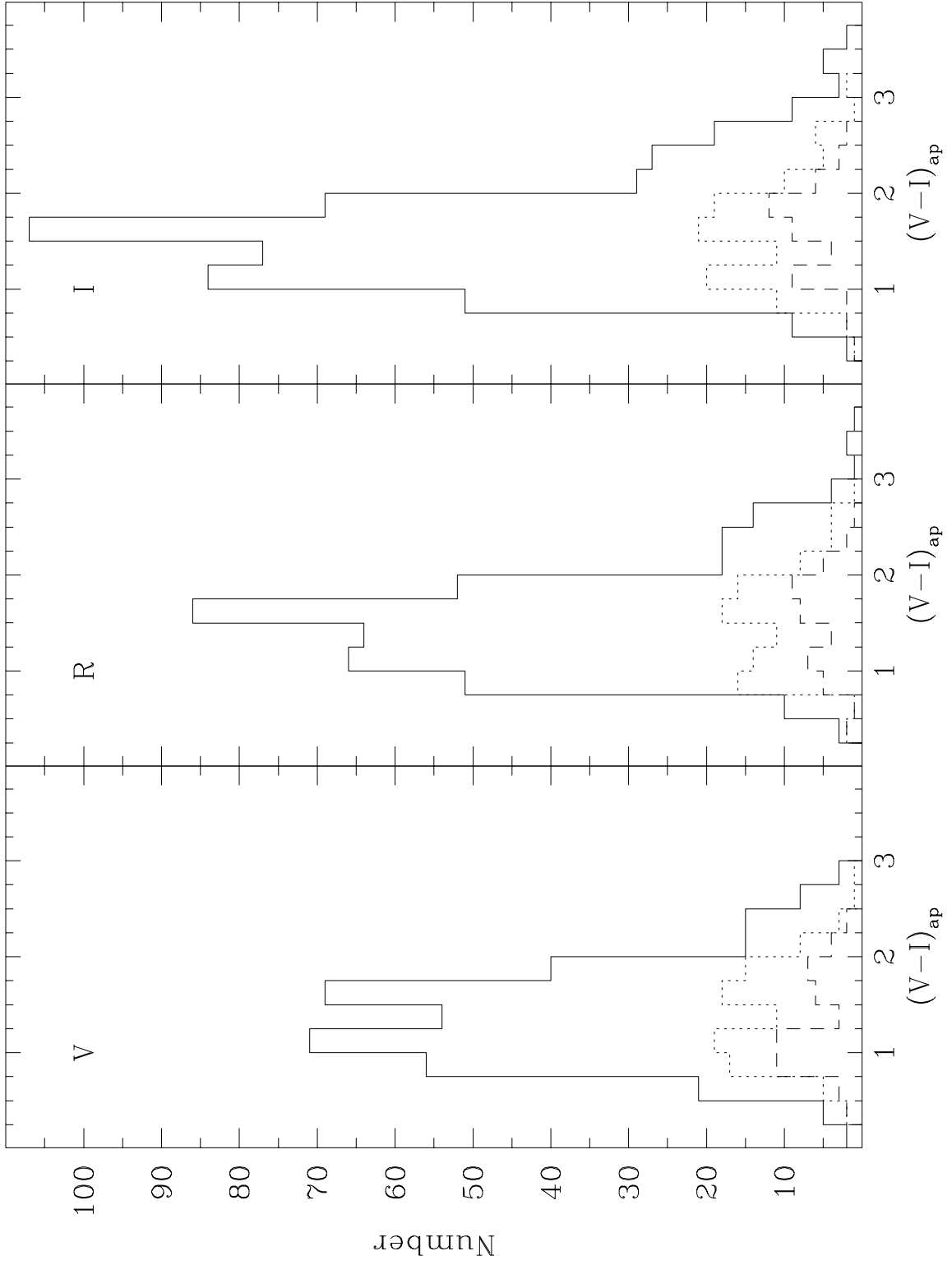